\documentclass[journal]{IEEEtran}
\ifCLASSINFOpdf
  \usepackage[pdftex]{graphicx}
  \graphicspath{{./figures/}}
\else
\fi
\usepackage{amsmath}
\usepackage{algorithmic}
\usepackage{algorithm}
\usepackage{multirow}
\usepackage{physics,bbm,mathtools,amsmath,amsthm,amssymb,amsfonts}
\usepackage{booktabs}
\usepackage{url}
\usepackage{xcolor}
\usepackage{cite}
\usepackage[pdftex]{graphicx}
\usepackage{amsmath}
\usepackage{array}

\newtheoremstyle{colon}
{}
{}
{\itshape}%bodyfont
{}%indent
{\bfseries}%headfont
{:}%head punctuation
{ }%space after head
{}
\theoremstyle{colon}
\newtheorem{proposition}{Proposition}[section]

\DeclareMathOperator*{\argmax}{arg\,max} % Jan Hlavacek

\begin{document}
\title{Unsupervised classification of the spectrogram zeros}


% \author{[Authors]~%\IEEEmembership{Member,~IEEE,}
%         Marcelo A. Colominas,~%\IEEEmembership{Fellow,~OSA,}
%         Nihls Laurent,~%\IEEEmembership{Life~Fellow,~IEEE}% <-this % stops a space
%         Sylvain Meignen,~%\IEEEmembership{Life~Fellow,~IEEE}% <-this % stops a space
%         François Auger,~%\IEEEmembership{Life~Fellow,~IEEE}% <-this % stops a space
% \thanks{Affiliations.}}

\author{Juan M. Miramont,~%\IEEEmembership{Member,~IEEE,}
        François Auger,~%\IEEEmembership{Fellow,~OSA,}
        Marcelo A. Colominas,~%\IEEEmembership{Life~Fellow,~IEEE}% <-this % stops a space
        Nils Laurent,~%\IEEEmembership{Life~Fellow,~IEEE}% <-this % stops a space
        and Sylvain Meignen~%\IEEEmembership{Life~Fellow,~IEEE}% <-this % stops a space
\thanks{Juan M. Miramont and François Auger are with Nantes Universit\'e, Institut de Recherche en \'Energie \'Electrique de Nantes Atlantique (IREENA, UR 4642), F-44600 Saint-Nazaire, France (email: juan.miramont@univ-nantes.fr and francois.auger@univ-nantes.fr).}% <-this % stops a space
\thanks{Marcelo A. Colominas is with the Institute of Research and Development in Bioengineering and \mbox{Bioinformatics} (IBB, UNER - CONICET), and with the Faculty of Engineering (UNER) Oro Verde, Entre Rios, Argentina (email: macolominas@conicet.gov.ar).}% <-this % stops a space
\thanks{Nils Laurent and Sylvain Meignen are with Jean Kuntzmann Laboratory, University of Grenoble-Alpes and CNRS UMR 5224, F-38401 Grenoble, France (email: nils.laurent1@univ-grenoble-alpes.fr and sylvain.meignen@univ-grenoble-alpes.fr).}%
\thanks{This work was supported by the ANR ASCETE project with grant number ANR-19-CE48-0001-01.}}


\markboth{ }%
{Shell \MakeLowercase{\textit{et al.}}: Bare Demo of IEEEtran.cls for IEEE Journals}



\maketitle

\begin{abstract}
The zeros of the spectrogram have proven to be a relevant feature to describe the time-frequency structure of a signal, originated by the destructive interference between components in the time-frequency plane. In this work, a classification of these zeros in three types is introduced, based on the nature of the components that interfere to produce them. Echoing noise-assisted methods, a classification algorithm is proposed based on the addition of independent noise realizations to build a 2D histogram describing the stability of zeros. Features extracted from this histogram are later used to classify the zeros using a non-supervised clusterization algorithm. A denoising approach based on the classification of the spectrogram zeros is also introduced. Examples of the classification of zeros are given for synthetic and real signals, as well as a performance comparison of the proposed denoising algorithm with another zero-based approach.
\end{abstract}

\begin{IEEEkeywords}
Zeros of the spectrogram, time-frequency \mbox{analysis}, non-stationary signals, noise-assisted methods.
\end{IEEEkeywords}

% \fi
\IEEEpeerreviewmaketitle
\section{Introduction}
% \makeatletter
% \def\convertto#1#2{\strip@pt\dimexpr #2*65536/\number\dimexpr 1#1}
% \makeatother
% \convertto{cm}{\the\textwidth}
Time-frequency (TF) representations, such as the widely-known \emph{spectrogram} \cite{flandrin1998time}, are powerful tools to reveal the time-varying frequency structure of a signal. A common interpretation of the spectrogram is that of a TF \emph{distribution} of the energy, a conception that naturally leads to consider that more information is located where the energy of the signal is more concentrated in the TF plane \cite{cohen1966generalized, auger1995improving,guillemain1996characterization,chassande1997differential, sejdic2009time}. %Consequently, a number of methods were built around the notion of \emph{high} values of the spectrogram. 

More recently, a paradigm-shifting idea was proposed in the work of Flandrin \cite{flandrin2015time}, where the zeros of the spectrogram have a central role instead of its larger values.
Indeed, it can be shown that zeros or \emph{silent points} of the spectrogram are relevant for the characterization of the TF structure of a signal \cite{gardner2006sparse, flandrin2015time, flandrin2016sound}.
In addition, Bardenet et al. \cite{bardenet2018zeros, bardenet2021time} showed that the distribution of the zeros of the spectrogram of complex circular white noise corresponds to that of the zeros of a planar Gaussian analytic function, establishing promising connections between signal processing and the study of spatial point processes.

The zeros of the spectrogram are created by \emph{destructive interference} between components in the short-time Fourier transform (STFT) \cite{flandrin2015time,flandrin2018explorations}, hence they are intimately related to the distribution of the signal energy in the plane. This will be a core idea throughout this work. We shall see that, when two signal components interfere, the position of the resulting zeros of the spectrogram seems to be robust to the addition of noise, for some low relative noise amplitudes with respect to the signal amplitude.

% In contrast, other critical points of the spectrogram seem to be less stable \cite{meignen2021interference}.Moreover, it has been empirically shown that the zeros of the spectrogram originated from the interference between signal components can be characterized by the energy surrounding them, among other criteria, allowing to discriminate them from the zeros located in noise only regions of the spectrogram \cite{meignen2021interference}.

This leads to the main three contributions of this paper. The first one is a criterion to classify the zeros of the spectrogram depending on the nature of the components that interfere to create them. We will show that it is possible to discriminate the zeros of the spectrogram in three categories: 1) the zeros created by the interference between deterministic signal components, 2) the zeros only related to noise components and 3) the zeros created through the interaction between signal and noise components. 

The second main contribution of this article is the introduction of 2D histograms that describe the spectrogram zeros variability when several independent noise realizations are added to the signal. We shall show that this representation can be used to correctly discriminate the zeros in the three categories described before by computing first a number of features from the 2D histogram and using a clusterization technique to classify them in a non-supervised, automatic way.

Finally, a third contribution is a denoising method based on the classification of the spectrogram zeros. We will describe a simple multicomponent signal for which other zero-based strategies face limitations, and illustrate how the here proposed approach can be more efficient for disentangling noise and signal in this case.

% Then we introduce a 2D histogram of the position of zeros, the study of which can lead to the automatic classification of the zeros in the categories described, 3) a noise-assisted algorithm to automatically classify the zeros of the spectrogram, 4) a denoising technique based on the classification.
% 

% We then introduce the second an algorithm for automatically classify the spectrogram zeros in the mentioned three kinds is described. This method is based on testing the stability of the zeros in the TF plane, by adding several independent noise realizations to the signal and computing a two-dimensional histogram that shows how the location of the zeros changed after the addition of noise. Finally, 

The rest of the paper is organized as follows. In Sec. \ref{sec::context} we define the STFT, the spectrogram and give other relevant elements to set the context of the article. In Sec. \ref{sec::three_types} we detail a criterion to classify the zeros in three types, whereas in Sec. \ref{sec::histograms} we introduce the 2D histograms of the position of zeros. Later, in Sec. \ref{sec::discriminating} we show how to automatically classify the zeros of the spectrogram using an unsupervised approach. In Sec. \ref{sec::results} we illustrate the performance of the classification algorithm with simulated and real signals, and we give an example of a denoising strategy derived from the previous findings. Finally, we comment on the obtained results in Sec. \ref{sec::discussion} and draw some conclusions in Sec. \ref{sec::conclusions}.

\section{Context} \label{sec::context}
Given a signal $x\in L^{1}\left( \mathbb{R} \right) \cap L^{2}\left( \mathbb{R} \right)$, we define its Fourier transform (FT) as
\begin{equation}\label{eq::ft}
\hat{x}(f) \coloneqq \int_{-\infty}^{+\infty} x(t)\; e^{-i2\pi f t} dt,
\end{equation}
where $t,f \in \mathbb{R}$ are the time and frequency variables. The STFT, which can be considered as a time-localized version of the Fourier transform, is a fundamental tool for time-frequency signal analysis. We define the STFT of $x(t)$ as 
\begin{equation}\label{eq::stft}
V_{x}^{g}(t,f) \coloneqq \int_{-\infty}^{+\infty} x(u)\; g(u-t)^{*}\; e^{-i2\pi f u} du,
\end{equation}
where $g\in L^{1}\left( \mathbb{R} \right) \cap L^{2}\left( \mathbb{R} \right)$a and $g(t)^{*}$ is complex conjugate of $g(t)$. Henceforth, we consider the analysis window $g(t)$ as a unitary Gaussian window given by
\begin{equation} \label{eq::window}
g(t) =  \frac{2^{1/4}}{\sqrt{T}} \;  e^{-\dfrac{\pi t^2}{T^2}}, 
\end{equation}
where $T$ determines its width.
We shall consider $T=1$ s, unless stated otherwise, so that the window has the same essential support in time and frequency.

The spectrogram of $x(t)$, defined as the squared modulus of the STFT, will be given then by
\begin{equation}
 S^{g}_{x}(t, f) \coloneqq \left| V^{g}_{x}(t, f) \right|^{2}.
\end{equation}

We define the discrete-time counterpart of Eq. \eqref{eq::stft} as
\begin{equation}\label{eq::discrete_stft}
 V_{x}^{g}[n,q] \coloneqq \sum_{m=n-L}^{n+L} x[m] g[m-n]e^{-\dfrac{i2\pi q m}{N}},
\end{equation}
where $n,q \in \mathbb{Z}$ are the discrete time and frequency variables, $L$ is the half-width of the discretized analysis window $g[n]$ and $N\in\mathbb{N}$ is the number of frequency bins of the discrete frequency axis.

In the following, we shall consider signal and noise mixtures. What one means by \mbox{\emph{signal}} and \emph{noise} depends on the application, but usually it can be considered that the signal is a quantity of interest that exhibits some \emph{organization} and, in contrast, noise is usually regarded as a \emph{random} fluctuation including all the other influences on the observed phenomena that are not of interest \cite{flandrin2018explorations}.

We will consider $\xi(t)$ a real white Gaussian noise realization, where $\xi(t)$ is distributed as $\mathcal{N}(0,\gamma_{0}^2)$ for all values of $t$, satisfying:
\begin{equation}
    \mathbb{E}\left\lbrace \xi(t)\xi(t-\tau) \right\rbrace  = \gamma_{0}^{2}\delta(\tau),
\end{equation}
where $\gamma_{0}^{2}$ is the noise variance and $\delta(t)$ is the Dirac distribution.
We shall express the Signal-to-Noise Ratio between a deterministic signal $x(t)$ and $\xi(t)$ as
\begin{equation}
 \operatorname{SNR}(x,\xi) = 10 \log_{10} \left( \frac{\norm{x}_{2}^{2}}{\gamma_{0}^{2}} \right) \;\text{(dB)}, \label{eq::snr}
\end{equation}
where $\norm{x}_{2}$ is the usual 2-norm of $x$ and its square is the energy of the signal.

\begin{figure}
\centering
 \setlength{\tabcolsep}{0pt}
 \renewcommand{\arraystretch}{0.0}
 \hspace*{-0.5cm}
 \begin{tabular}{cc}
  \includegraphics[width = 0.26\textwidth]{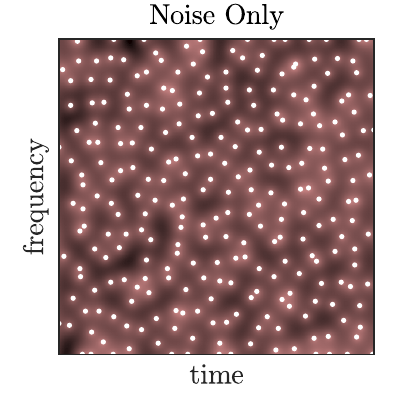} &
  \includegraphics[width = 0.26\textwidth]{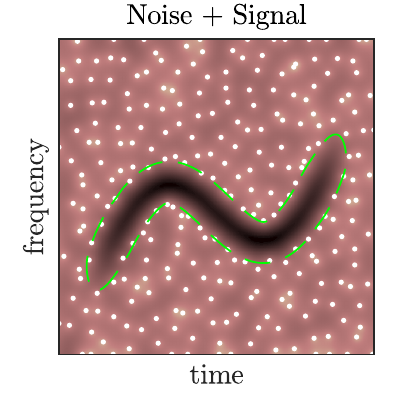} \\
  (a) & (b) \\
 \end{tabular}
 \caption{(a) Logarithm of the spectrogram of real white Gaussian noise. (b) Spectrogram of the mixture of the same noise realization with a signal. The dashed line indicates the curve $\Gamma$ given by Eq. \eqref{eq::level_set}. In both cases, the superimposed white dots mark the positions of the spectrogram zeros. }\label{fig::level_set}

\end{figure}

\section{Three Types of Zeros of the Spectrogram} \label{sec::three_types}
The zeros of the spectrogram are a consequence of the \emph{destructive interference} between components in the time-frequency plane \cite{flandrin2015time, flandrin2018explorations}. Indeed, the spectrogram of the sum of two signals $x_{1}(t)$ and $x_{2}(t)$ can be written as \cite{kadambe1992comparison, jeong1992mechanism}
\begin{multline}\label{eq::spect_mixture}
 S^{g}_{x_{1}+x_{2}}(t,f) = S^{g}_{x_{1}}(t,f) + S^{g}_{x_{2}}(t,f) \\ +  2 \real\{V^{g}_{x_{2}}(t, f){V^{g}_{x_{1}}}(t, f)^{*}\}.
\end{multline}
Expressing $V^{g}_{x_{1}}(t, f) = M_{x_{1}}(t,f)\;e^{i\Phi_{x_{1}}(t,f)}$ and $V^{g}_{x_{2}}(t, f) = M_{x_{2}}(t,f)\;e^{i\Phi_{x_{2}}(t,f)}$, one obtains:
\begin{multline}\label{eq::spect_mixture_polar}
 S^{g}_{x_{1}+x_{2}}(t,f) = M_{x_{1}}(t,f)^{2} + M_{x_{2}}(t,f)^{2} \\ + 2 M_{x_{1}}(t,f)M_{x_{2}}(t,f) \cos\left(\Phi_{x_{2}}(t,f) - \Phi_{x_{1}}(t,f) \right),
\end{multline}
from which the following proposition is derived:
\begin{proposition}\label{prop::conditions}
Destructive Interference. For $S^{g}_{x_{1}+x_{2}}(t,f)$ to be equal to zero at a point $(t,f)$, the following necessary and sufficient conditions must be satisfied:
\begin{enumerate}
 \item The phases $\Phi_{x_{1}}(t,f)$ and $\Phi_{x_{2}}(t,f)$ must differ from an integer odd multiple of $\pi$.
 \item The modulus of $V^{g}_{x_{2}}(t, f)$ and $V^{g}_{x_{1}}(t, f)$ must be equal.
\end{enumerate}
\end{proposition}
A proof of this simple proposition is given in the Appendix.
In the noiseless case, those conditions are fulfilled for some $(t,f)$ provided that the signal is multicomponent.
One can then consider $x_{1}(t)$ and $x_{2}(t)$ as two signal components that are linearly combined to produce the signal.
Thus, the zeros of the spectrogram would be generated by the interference between $x_{1}(t)$ and $x_{2}(t)$, and therefore they will be completely \emph{deterministic}.
We will denote them as zeros of the \emph{first} kind.

In the noise only case, shown in Fig. \ref{fig::level_set}a, the zeros are uniformly distributed in the TF plane, generated by the interference of multiple randomly located \emph{logons} (albeit not at arbitrary positions because of the restrictions posed by the reproducing kernel of the STFT \cite{flandrin2015time, flandrin2018explorations}).
Hence, $x_{1}(t)$ and $x_{2}(t)$ can be thought of as two adjacent logons, the interference of which would produce zeros in the spectrogram.
This case has been by far the most studied one, and the characterization of the zeros of noise components and the lattice-like structure they generate in the TF plane is the basis of some recently developed denoising methods \cite{flandrin2015time, bardenet2018zeros, ghosh2022signal}.
In the sequel, we will term the zeros created by the interference between noise logons, zeros of the \emph{second} kind.

Regarding the mixture of signal and noise, the situation is more complex. Comparing Fig. \ref{fig::level_set}a and \ref{fig::level_set}b, it is possible to see that the presence of the signal only affects the spectrogram zeros \emph{locally} \cite{flandrin2015time, bardenet2018zeros}, since zeros far away from the signal energy are not modified.  
Considering this, one can conclude that in the regions of the TF plane where the noise dominates, the zeros of the spectrogram will correspond to zeros of the second kind. In contrast, zeros of the first kind will be found in the regions where the signal components are more energetic than the noise, as described before, provided that the signal components are close enough.

It remains to analyze the zeros created by the interference between the deterministic signal components and the noise where neither of these influences can be neglected.
As expected, these zeros will be located close to the border of the signal domain.
Considering now $x_{1}(t)$ as a signal component and $x_{2}(t)$ as noise, it is not possible to give an exact description of the position of the zeros created by the interference between them because of the random nature of the noise. But despite this, one can approximate where the zeros will be located by replacing the spectrogram of noise in Condition 2 of Prop. \ref{prop::conditions} by its expected value (constant and proportional to the noise variance in the case of white Gaussian noise and a unitary energy window) \cite{flandrin1998time, flandrin2018explorations}.
Then, Condition 2 can be written as
\begin{equation} \label{eq::level_set}
 S^{g}_{x_{1}}(t,f)  =  \gamma_{0}^{2}.
\end{equation}
Equation \eqref{eq::level_set} defines a level curve $\Gamma$ given by
\begin{equation}
\Gamma = \left\lbrace (t,f) \;|\; S^{g}_{x_{1}}(t,f)  =  \gamma_{0}^{2} \right\rbrace,
\end{equation}
that well approximates the location of the zeros that surround the signal domain, as illustrated by Fig. \ref{fig::level_set}b, where the dashed line indicates the level curve $\Gamma$.
This poses a restriction on the zeros created by the interference of signal and noise, in the sense that they are created in the neighborhood of $\Gamma$, making these zeros of a very different nature from the first two categories introduced before.
Hence we will consider these zeros as belonging to a \emph{third} kind: those that are created by the interference between signal and noise components.

For lower (resp. higher) values of $\operatorname{SNR}(x,\xi)$, the domain determined by the interior of $\Gamma$ progressively shrinks (resp. expands), and constitutes a good approximation to the signal domain \cite{ghosh2022signal}.
Notice that, however, one usually does not have access to $S^{g}_{x_{1}}(t,f)$ and an estimation of $\gamma_{0}$ is commonly used for denoising by thresholding $S_{x+\xi}(t,f)$ \cite{donoho1994ideal, mallat2008wavelet, pham2018novel}.

\begin{figure}
\centering
 \setlength{\tabcolsep}{0pt}
 \renewcommand{\arraystretch}{0.0}
%  \hspace*{-0.7cm}
 \begin{tabular}{c}
  \includegraphics{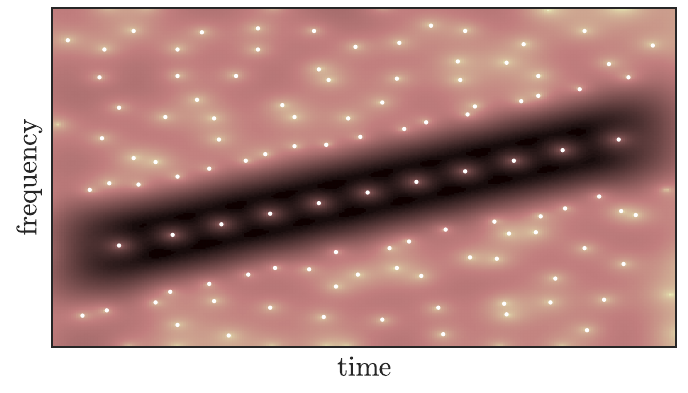} \\
  (a) \\
  \includegraphics{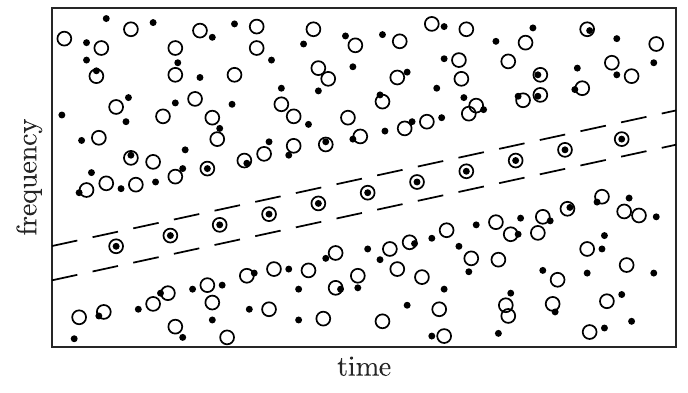} \\
  (b) \\
 \end{tabular}
 \caption{(a) Logarithm of the spectrogram of a signal with two parallel linear chirps contaminated by real white Gaussian noise. (b) Position of the spectrogram zeros. Circles and dots indicate the position of the spectrogram zeros corresponding to two different noise realizations. Dashed lines denote the instantaneous frequency associated with each chirp.}\label{fig::position_of_zeros}
\end{figure}

\begin{figure*}
\centering
 \setlength{\tabcolsep}{0pt}
 \renewcommand{\arraystretch}{0.5}
\begin{tabular}{ccc}
 \includegraphics{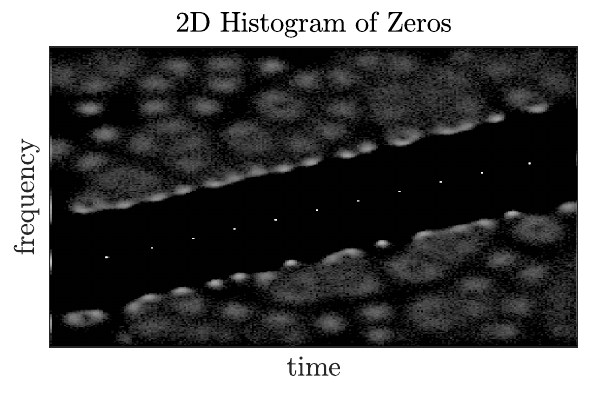}&
 \includegraphics{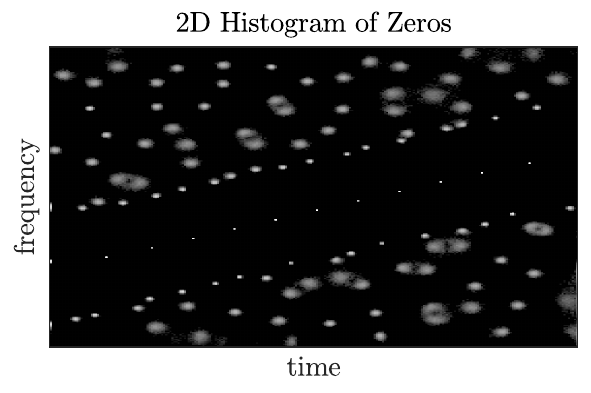} &
 \includegraphics{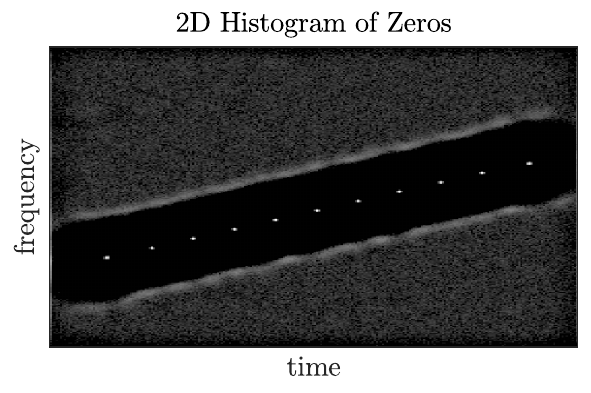} \\
 (a) & (b) & (c) \\
\end{tabular}
 \caption{ 2D histograms of new zeros after noise perturbation corresponding to the signal shown in Fig. \ref{fig::position_of_zeros}a. (a) Added noise realizations with the same variance as the original one, $\gamma^{2}_{j} = \gamma^{2}_{0}$. (b) Noise realizations with a lower variance, $\gamma^{2}_{j} = \frac{1}{20}\;\gamma^{2}_{0}$. (c) Noise realizations with higher variance $\gamma^{2}_{j} = 20 \; \gamma^{2}_{0}$. }\label{fig::spectrogram_and_histogram}
\end{figure*}

\section{A 2D Histogram of the Position of Zeros} \label{sec::histograms}

\subsection{A noise-assisted Study of Zeros}
Consider a signal composed of two parallel linear chirps and contaminated with real white Gaussian noise, the spectrogram of which is shown in Fig. \ref{fig::position_of_zeros}a.

Hypothetically, if one changed the noise realization of the signal shown in Fig. \ref{fig::position_of_zeros}a for another one with the same variance, and then proceeded to compare the position of the zeros of the spectrogram of the original noisy signal with the zeros of the modified one, it would appear as if the zeros have \emph{moved}. This effect can be seen in Fig. \ref{fig::position_of_zeros}b, where the circles and dots indicate, respectively, the position of the spectrogram zeros with the original noise realization and a new one.

One can notice that the zeros between the chirps, i.e. zeros of the first kind, are not affected by this change. This is somehow expected, because of the predominantly deterministic nature of these zeros. However, this is not true for the zeros of the second and third kind. As one can observe in Fig. \ref{fig::position_of_zeros}b, the new zeros are created at different locations in the remaining regions of the TF plane when the noise realization is modified.

Because the new noise realization has the same variance as the original one, the signal domain determined by $\Gamma$ in Eq. \eqref{eq::level_set} is not affected. This may explain why Fig. \ref{fig::position_of_zeros}b also shows that the new zeros created near the border of the signal domain seem to be located closer to the original zeros. This effect is not observed in the noise-dominated regions of the plane, where the locations of the new zeros look less correlated to that of the original ones. These observations lead to a principle that could be used to differentiate between the different types of zeros by studying \emph{how far} the new zeros are created from the original ones after modifying the noise.

Nevertheless, changing only the noise component of a signal for an equivalent one is not feasible in practice. Drawing inspiration from the noise-assisted methods \cite{wu2009ensemble, torres2011complete}, one can rather \emph{add} a new realization of noise that will then modify the position of all the zeros of the spectrogram.
To see why this occurs, let us first call $y(t) = x(t) + \xi(t)$ the \emph{original} signal mixture, where $\xi(t)$ is the noise already present in the signal, and $y_{j}(t) = x(t) + \xi(t) + \eta_{j}(t)$ a \emph{new} mixture, where $\eta_{j}(t) \;j\in\mathbb{N}$ is a noise realization (here real white Gaussian noise) with zero mean and variance $\gamma_{j}^{2}$. 

Then, the STFT of $y_{j}(t)$ can be written as the following series (see \cite{grochenig2001foundations}, chapter 3, and \cite{bardenet2018zeros} for more details):
\begin{equation}
 V_{y_{j}}^{g}(t, f) = e^{i \pi f t } e^{-\pi\abs{z}^{2}/2} \sum_{k=0}^{\infty} \left\langle y_{j}, h_{k} \right\rangle \frac{\pi^{k/2}z^{k}}{\sqrt{k!}} \\ \label{eq::planar_gaf}                            
\end{equation}
where $z = t+if$, $\{h_{k}\}_{k=0}^{\infty}$ is the set of Hermite functions and $\left\langle y_{j}, h_{k} \right\rangle$ is the inner product of $y_{j}(t)$ and $h_{k}(t)$. Notice that the zeros of $V_{y_{j}}^{g}(t, f)$ are then determined by the coefficients $\left\langle y_{j}, h_{k} \right\rangle$. By linearity, we can rewrite Eq. \eqref{eq::planar_gaf} as
\begin{equation}
V_{y_{j}}^{g}(t, f) = e^{i \pi f t } e^{-\pi\abs{z}^{2}/2} \sum_{k=0}^{\infty} (a_{k} + b_{j,k}) \frac{\pi^{k/2}z^{k}}{\sqrt{k!}} \label{eq::coef_perturbation} 
\end{equation}
where $a_{k} = \left\langle x+\xi ,h_{k}\right\rangle$ and $b_{j,k} = \left\langle \eta_{j} ,h_{k}\right\rangle$.
Because the coefficients $a_{k}$ determine the zeros of $S_{y}^{g}(t, f)$, i.e. the zeros of the spectrogram of the original mixture, one can consider $b_{j,k}$ as a random \emph{perturbation} of $a_{k}$, distributed as $\mathcal{N}(0,\gamma_{j}^{2})$ \cite{holden1996stochastic,bardenet2018zeros}, the effect of which is changing the location of the zeros of $S_{y}^{g}(t, f)$.

If the variance of the noise is negligible, then the coefficients $a_{k}$ are barely perturbated, producing no appreciable changes in the positions of the zeros of $S_{y}^{g}(t, f)$. On the contrary, if $\gamma_{j}^{2}$ is very large, the original coefficients $a_{k}$ become negligible instead, and each realization produces an almost completely different set of zeros.

\subsection{Computing 2D Histograms of the Position of Zeros}
Since mapping the coefficients of the series in Eq. \eqref{eq::coef_perturbation} to the positions of the zeros is not feasible, the preceding analysis is not useful \emph{per se} to discriminate between the types of zeros described in Sec. \ref{sec::three_types}.
However, it provides the basic idea to develop an heuristic for the classification of the spectrogram zeros based on adding new realizations of noise $\eta_{j}(t)$ to the signal and then finding the zeros of the spectrogram of these new mixtures.
After repeating this procedure for $J$ independent noise realizations, a 2D histogram can obtained by counting the number of zeros that have fallen in each TF position throughout the ensemble of realizations. An example of this histogram for the signal shown in Fig. \ref{fig::position_of_zeros} is depicted in Fig. \ref{fig::spectrogram_and_histogram}a. Such an histogram can be regarded as an indirect observation of the perturbation $b_{j,k}$ in Eq. \eqref{eq::coef_perturbation}, induced by the addition of noise.

In practice, one can at most approximate the position of the zeros as minima on the grid of TF points produced by discretization of the STFT given by Eq. \eqref{eq::discrete_stft} \cite{flandrin2015time,bardenet2018zeros,escudero2021efficient}. Consequently, the binning of the aforementioned 2D histogram will correspond to that of the discrete TF plane. Throughout this work, zeros are found as the local minimum on a \mbox{$3\times3$} grid centered at each point of the (discrete) spectrogram, which is well justified by the properties of complex analytic functions \cite{boas2011entire}. On this regard, note that efficient location of the spectrogram zeros is a topic of current research \cite{escudero2021efficient}.

Algorithm \ref{alg::generating_2Dhist} summarizes the steps used here to generate a 2D histogram of the position of zeros of the spectrogram. The algorithm receives the number of noise realizations to use in order to compute the histogram, and the variance of the noise realizations. We shall denote by $\mathcal{Z}_{0}$ the set of zeros of the spectrogram of $y(t)$, the \emph{original} zeros, as opposed to the sets $\mathcal{Z}_{j}$ of \emph{new} zeros of the spectrogram of each $y_{j}(t),\; j=1,2,...J$. Then, each time a new mixture $y_{j}(t) = y(t) + \eta_{j}(t)$ is generated, the set of new zeros $\mathcal{Z}_{j}$ is computed and the histogram $G[n,q]$ is updated by adding $1$ to each position $[n,q]\in \mathcal{Z}_{j}$.

\begin{algorithm}[b]
\begin{algorithmic}[1]
\REQUIRE The number of realizations $J$, the variance of the noise realizations $\gamma_{\operatorname{noise}}^{2}$.
\STATE Initialize $G[n,q] = 0, \; \forall \; n,q $.
\FOR{$1\leq j\leq J$} 
\STATE Generate a new realization of noise $\eta_{j}(t)$, with variance $\gamma^{2}_{j} = \gamma^{2}_{\operatorname{noise}}$.
\STATE Generate a mixture $y_{j}(t) = y(t) + \eta_{j}(t)$.
\STATE Compute the (discrete) spectrogram $S_{y_{j}}[n,q]$ of $y_{j}(t)$.
\STATE Find the set of zeros $\mathcal{Z}_{j}$ of $S_{y_{j}}[n,q]$.
\STATE Update the histogram as $G[n,q] = G[n,q] + \mathbbm{1}_{\mathcal{Z}_{j}}[n,q]$, where $\mathbbm{1}_{\mathcal{Z}_{j}}[n,q] = 1$ if $[n,q]\in \mathcal{Z}_{j}$, and $\mathbbm{1}_{\mathcal{Z}_{j}}[n,q] = 0$ otherwise.
\ENDFOR
\RETURN $G[n,q]$.
\end{algorithmic}
\caption{2D histogram of the positions of zeros} \label{alg::generating_2Dhist}
\end{algorithm}

\subsection{Local Density and Concentration of the 2D Histograms}
Figs. \ref{fig::spectrogram_and_histogram}a to \ref{fig::spectrogram_and_histogram}c show the 2D histograms corresponding to different variances of the added noise, for $J = 512$ and $\operatorname{SNR}(x,\xi) = 30$ dB. The brighter the histogram, the higher the number of zeros that have fallen in that particular point. Considering the number of zeros located within a neighborhood of an original zero as a local \emph{density} of the histogram, one can see that the histogram is denser, i.e. brighter, around the positions of the original zeros. Particularly, one can observe a higher density of new zeros at the border of the signal domain, which is clearly highlighted in the histograms shown in Fig. \ref{fig::spectrogram_and_histogram}, and at the zeros between the signal components.

As is also shown in Fig. \ref{fig::position_of_zeros}b, the zeros created in the region between the modes are the most \emph{stable} ones, meaning that they are barely affected by the noise added to the signal, provided that the variance of the added noise realizations is not too high. A consequence of this is that new zeros are located exactly in the same position as, or slightly dispersed around, the original zeros of the first kind.

Although the number of new zeros fallen in the neighborhoods of two original zeros might be the same, they could be distributed very differently within this region. For example, they could be uniformly distributed or they could all fall at a unique position (like the zeros between the components in Fig. \ref{fig::position_of_zeros}b). Therefore, how the density of the histogram is distributed within the region surrounding an original zero, i.e. its local \emph{concentration}, is another aspect to describe the different kinds of zeros.

Contrary to what happens with the zeros between the signal components, the histogram concentration around the zeros located in the noise-only regions of the TF plane seems to be lower, meaning that new zeros appear at locations that are farther from the closest original zero and are more uniformly distributed within the neighborhood. An intermediate situation can be observed regarding the original zeros at the border of the signal domain: new zeros appear neither very close nor very far from them. Consequently, one can consider that the zeros located at the border of the signal domain are more stable to the addition of noise than those in the noise-only regions, but less stable than those generated by the interference of the signal with itself.

From Fig. \ref{fig::spectrogram_and_histogram}b it is possible to see that adding noise with a much lower variance does not allow to discriminate between the types of zeros, as the concentration of the histogram around those located at the border of the signal domain is very similar to that of the zeros located in the noise-only regions. The same happens if too much noise is added. This latter case is represented in Fig. \ref{fig::spectrogram_and_histogram}c, where one notices that the zeros of the second and third kind seem indistinguishable.

\section{Discriminating Between Zeros} \label{sec::discriminating}
We now describe how to automatically classify the zeros in the three types described in Sec. \ref{sec::three_types}. First, we define an estimator of the variance of the noise in the signal, which will be used to compute the aforementioned 2D histogram of the position of zeros. Then, we explain how the original zeros can be characterized by local descriptors of the density and concentration of this histogram. Finally we show how to classify the zeros by means of a clusterization algorithm to automatically group the zeros with similar characteristics.   

\subsection{Noise Variance Estimation}
In order to produce a meaningful histogram, it is needed to add independent noise realizations with the same variance as the original noise present in the signal (like in Fig. \ref{fig::spectrogram_and_histogram}a).
We use here the robust median absolute deviation estimator \cite{donoho1994ideal, mallat2008wavelet, pham2018novel} given by
\begin{equation} \label{eq::mad_estimator}
 \hat{\gamma}_{0} = \frac{\sqrt{2}}{0.6745} \; \operatorname{median}\left( \left| \real\left\lbrace V^{g}_{y}(t,f) \right\rbrace \right| \right),
\end{equation}
to estimate the standard deviation of the noise already present in the signal, the square of which can be then used as an input parameter in Algorithm \ref{alg::generating_2Dhist}.

\subsection{Local Density and Concentration Descriptors}
Based on the observations made in Sec. \ref{sec::histograms}, we will locally assess the distribution of the 2D histogram, i.e. in the neighborhood of each original zero, with the purpose of discriminating among the three described categories of zeros.

Such a neighborhood can be formalized as a ball centered at an original zero, with a radius $r\in\mathbb{R}^{+}$.
Then, the values of the 2D histogram in this ball are given by
\begin{equation} \label{eq::vicinity}
 B({\bf{z}},r) = \{G[n,q]\; |\; \operatorname{d}\left([n,q], \bf{z} \right) < r\}, \bf{z}\in\mathcal{Z}_{0},
 \end{equation}
where $\bf{z}$ is the center of the ball and represents the two coordinates of an original zero, and $\operatorname{d}([n,q],\bf{z})$ is the Euclidean distance between $\bf{z}$ and the point with coordinates $[n,q]$. One possible local measure of the density of the histogram can be then defined by simply summing the values inside the ball $B({\bf{z}},r)$, which is equivalent to $\Vert B({\bf{z}},r)\Vert_{1}$, i.e. the 1-norm of $B({\bf{z}},r)$.
As remarked in the previous section, we also need to take into account the local concentration of the histogram. Therefore, we combine $\Vert B({\bf{z}},r)\Vert_{1}$ with a measure of density concentration like the so-called Rényi entropy \cite{baraniuk2001measuring, meignen2017demodulation}, given by
\begin{equation} \label{eq::renyiH}
 H_{\alpha}(B({\bf{z}},r)) = \frac{\alpha}{1-\alpha} \log_{2} \left(  \norm{B_{\text{n}}({\bf{z}},r)}_{\alpha} \right),
\end{equation}
where $B_{\text{n}}({\bf{z}},r) = B({\bf{z}},r)/\Vert B({\bf{z}},r)\Vert_{1} $. Of course, this measure is computed provided that $\Vert B({\bf{z}},r)\Vert_{1} \neq 0 $.

Two values of $\alpha$ were studied, $\alpha\rightarrow \infty$, which results in the 
\emph{min-entropy} $H_{\infty}(B({\bf{z}},r))$, and $\alpha\rightarrow 1$, which gives the known Shannon entropy $H_{1}(B({\bf{z}},r))$.
Best results where found using the latter.

Notice that the higher the concentration of the histogram around the point ${\bf{z}}$, the lower will be the value of $H_{1}(B({\bf{z}},r))$. Zeros of the first kind are expected to have both high density, i.e. high $\Vert B({\bf{z}},r)\Vert_{1}$, and high concentration, i.e. low $H_{1}(B({\bf{z}},r))$. On the other hand, zeros of the second kind, as shown in Fig. \ref{fig::spectrogram_and_histogram}a, should have both low density and concentration. Zeros of the third kind represent an intermediate situation, with higher density than second kind zeros, but lower concentration than first kind ones.  

\begin{figure}
\centering
\includegraphics{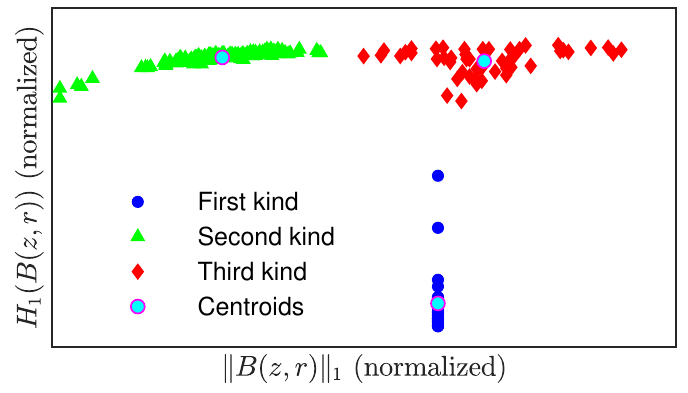}
\caption{Feature space of $\Vert B({\bf{z}},r)\Vert_{1}$ and $H_{1}(B({\bf{z}},r))$, obtained with the parameters described in Sec. \ref{sec::results::parameters}. Each point represents one zero of the spectrogram of the parallel chirps signal displayed in Fig. \ref{fig::zeros_classification}a.} \label{fig::clusters}
\end{figure}

\subsection{Zeros Clustering and Labeling}
Based on the previous observations, one can expect zeros of the same kind to aggregate and form clusters in the bidimensional space spanned by $\Vert B({\bf{z}},r)\Vert_{1}$ and $H_{1}(B({\bf{z}},r))$. Then, a clustering algorithm such as the widely known $K$-means \cite{bishop2006pattern}, can be applied to identify them. This algorithm looks for a prescribed number of groups $K\in\mathbb{N}$, given as a parameter. However, because one cannot guarantee that there will be always zeros of the three kinds, it is first necessary to decide whether $K=1$ (only noise is present in the signal), $K=2$ (there is a noisy signal and the spectrogram does not have zeros of the first kind) or $K=3$ (all three kinds of zeros are present). To make this decision automatic, one has first to apply the clustering algorithm with $K=1$, $K=2$ and $K=3$. Then, the number of clusters can be estimated by optimizing an appropriate measure, such as the GAP criterion \cite{tibshirani2001estimating}. 

Following the identification of the clusters, each one needs to be labeled in order to finally classify the zeros. Here we use $\Vert B({\bf{z}},r)\Vert_{1}$ and $H_{1}(B({\bf{z}},r))$ to assign a label to each of the clusters, following the example given in Fig. \ref{fig::clusters}. 
If all zeros are of the same type, i.e. $K=1$, then all the zeros are of the second kind, meaning that only noise is present in the signal.
When $K=2$ or $K=3$ we first identify the zeros of the second kind as those corresponding to the cluster whose centroid has the lowest value of $\Vert B({\bf{z}},r)\Vert_{1}$. Then, if $K=2$, the remaining cluster corresponds to the zeros of the third kind. On the other hand, if $K=3$,  two clusters still remain to be labeled. From these, the one whose centroid has the lowest value of $H_{1}(B({\bf{z}},r))$ corresponds to the zeros of the first kind, while the remaining cluster is finally  labeled as the one corresponding to the zeros of the third kind.

Algorithm \ref{alg::discriminating_zeros} summarizes the steps taken for the discrimination of the three types of zeros. The input parameters are the number $J$ of noise realizations used to compute the 2D histogram and the radius $r$ of the vicinity of zeros. These parameters were chosen by maximizing the accuracy of the discrimination of zeros of the spectrogram. The description of these experiments and the chosen values for the parameters are given in Sec. \ref{sec::results::parameters}.
\begin{algorithm}
\begin{algorithmic}[1]
\REQUIRE Number of realizations $J$ and the radius $r$ defining a neighborhood around zeros.
\STATE Estimate the variance of the noise $\hat{\gamma}_{0}$ using Eq. \eqref{eq::mad_estimator}.
\STATE Compute the histogram of the position of new zeros, $G[n,q]$, using Algorithm \ref{alg::generating_2Dhist}, with $J$ and $\gamma_{\operatorname{noise}} = \hat{\gamma}_{0}$.
\FOR{each $\bf{z} \in \mathcal{Z}_{0}$}
\STATE Determine the neighborhood $B({\bf{z}},r)$ of $\bf{z}$ as in Eq. \eqref{eq::vicinity}.
\STATE Compute $\Vert B({\bf{z}},r)\Vert_{1}$. 
\STATE Compute $H_{1}\left(  B({\bf{z}},r) \right)$ given by Eq. \eqref{eq::renyiH}.
\ENDFOR
\STATE Normalize the values of $\Vert B({\bf{z}},r)\Vert_{1}$ and $H_{1}\left(  B({\bf{z}},r) \right)$ to have zero mean and unitary variance.
\STATE Apply $K$-means algorithm with $K=1,2,3$ to the normalized descriptors using \emph{cityblock} distance.
\STATE Decide whether $K=1$, $K=2$ or $K=3$ by means of the GAP criterion \cite{tibshirani2001estimating}.

\IF{K=1}
\STATE The mixture is only noise, label all the zeros as zeros of the second kind.
\ELSE
\STATE Label the zeros belonging to the cluster whose centroid has the lowest value of $\Vert B({\bf{z}},r)\Vert_{1}$ as zeros of the second kind.
\IF{K=3}
\STATE Consider the remaining two unlabeled clusters. Label the zeros belonging to the cluster whose centroid has the lowest value of $H_{1}\left( B({\bf{z}},r) \right)$ as zeros of the first kind.
\ENDIF
\STATE Label the zeros of the remaining cluster as zeros of the third kind.
\ENDIF
\RETURN Labels of each zero.
\end{algorithmic}
\caption{Classification of zeros of the spectrogram.} \label{alg::discriminating_zeros}
\end{algorithm}

\subsection{TF Filtering Based on Classified Zeros} \label{sec::tf_filtering}
In \cite{flandrin2015time, flandrin2016sound}, a strategy for signal and noise disentanglement is proposed based on the Delaunay triangulation of the zeros of the spectrogram. The rationale behind this method is that the presence of a signal creates larger-than-expected regions without zeros in the TF plane when compared to the noise-only case. Consequently, triangles with at least one edge longer than some threshold $\ell_{\max}$ are selected as those corresponding to the signal domain. The value of $\ell_{\max}$ must be chosen to minimize the probability of selecting noise triangles, for example $\ell_{\max}=2.0$ as in \cite{flandrin2015time}, so that mostly triangles corresponding to the signal domain are selected. Such a procedure works robustly for monocomponent signals, but some limitations for multicomponent signals might appear. We shall now describe an example that illustrates such limitations.

Consider a simple signal with three tones at frequencies $f_{1}$, $f_{2}$ and $f_{3}$, such as
\begin{equation} \label{eq::triple_tone}
 x_{h}(t) = s_{f_{1}}(t) + s_{f_{2}}(t) + s_{f_{3}}(t),
\end{equation}
where $s_{f_{i}}(t) = \cos(2\pi f_{i}t)$, and $f_{1}<f_{2}<f_{3}$.

Due to the presence of multiple components, deterministic zeros of the spectrogram will appear between them, where the conditions given in Prop. \ref{prop::conditions} are met. Linearity property allows one to write the STFT of $x_{h}(t)$ as the sum of the STFT of each individual tone, given by
\begin{equation}
 V_{s_{f_i}}^{g}(t,f) = \frac{1}{2}\hat{g}(f-f_{i})e^{-i2\pi (f-f_{i})t} + \frac{1}{2} \hat{g}(f+f_{i})e^{-i2\pi (f+f_{i})t}.
\end{equation}

\begin{figure}
\centering
 \setlength{\tabcolsep}{0pt}
 \renewcommand{\arraystretch}{0.5}
 \begin{tabular}{c}
  \includegraphics{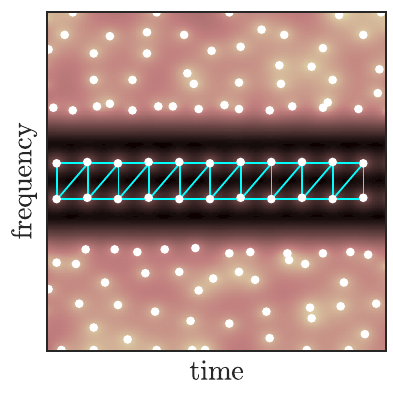} 
%   \includegraphics[width = 0.12\textwidth]{parallel_tones_det_tri} \\
%  (a)& (b) \\
\end{tabular}
\caption{The logarithm of the spectrogram of a signal with three Gaussian impulsive transients defined in Eq. \eqref{eq::triple_tone}, with $\operatorname{SNR}(x_{h},\xi) = 30$ dB. White dots indicate the spectrogram zeros. Lines indicate the Delaunay triangles, the vertices of which are the spectrogram zeros between adjacent components. }\label{fig::tones}
\end{figure}

Considering only the positive frequencies due to the symmetry of the STFT, let us define the differences between the frequency location of adjacent tones as $\Delta f_{1,2} = f_{2}-f_{1}$ and $\Delta f_{2,3} = f_{3}-f_{2}$, and consider the following condition on the separation of the tones:
\begin{equation}
\Delta f_{1,2} = \Delta f_{2,3} =  \frac{\theta}{\sqrt{2\pi}},
\end{equation}
where $\theta>0$ and $\frac{1}{\sqrt{2\pi}}$ is the standard deviation of the window $\hat{g}(f)$ defined by Eq. \eqref{eq::window} for $T=1$ s. 
If one chooses $\theta\geq3$, then this condition means that their STFTs are separated by at least three times the standard deviation of $\hat{g}(t)$. As a consequence, the interaction between the STFTs of non-adjacent tones, i.e. between the first and the third, can be neglected. This simplification allows one to compute the location of the zeros of the spectrogram by considering only contiguous tones. Bearing this in mind, one can find that the deterministic zeros will be located exactly between each pair of tones, at locations $\tilde{f}_{1,2} = (f_{1}+f_{2})/2$ and $\tilde{f}_{2,3} = (f_{2}+f_{3})/2$. Meanwhile, the time coordinates of the zeros between the first and second tone are given by
\begin{equation}
 \tilde{t_{k}} = \frac{k+1/2}{f_{2}-f_{1}} = \frac{k+1/2}{\Delta f_{1,2}},
\end{equation}
where $k\in\mathbb{Z}$. Notice that if $\Delta f_{1,2}=\Delta f_{2,3}$, then the zeros located between the first and second tones have the same time coordinates as the ones located between the second and the third. This situation is illustrated in Fig. \ref{fig::tones}, which shows that the Delaunay triangles obtained in this case are, ideally, orthogonal triangles. Hence, one can compute a good approximation of the length of their longest edge as
\begin{equation} \label{eq::sep_condition}
 \ell = \sqrt{\frac{(2\pi)^{2} + \theta^{4}}{2\pi\theta^{2}}}. 
\end{equation}

Now, taking $\theta=3$, the length of the longest edge of these triangles is approximately $\ell \approx 1.46$. This means that, in order to be able to select these triangles as part of the signal domain, a threshold $\ell_{\max}<1.46$ must be used in the algorithm proposed by Flandrin in \cite{flandrin2015time}. Nevertheless, this value is too low to allow a complete disentanglement between signal and noise because triangles corresponding to noise regions are very likely to be selected (see Fig. 2 from \cite{flandrin2015time}).

This constitutes an example in which the classification of zeros in three types can be useful to determine the signal domain, by selecting the Delaunay triangles that satisfy any of the following two criteria:
\begin{enumerate}
 \item At least one of their vertices is a zero of the first kind (i.e. zeros created by interference between the components of the signal).
 \item All their vertices are zeros of the third kind (i.e. zeros located at the border of the signal domain).
\end{enumerate}

The first criteria aims to keep the Delaunay triangles associated with zeros of the first kind, since these zeros are surrounded with high energy from the signal. Meanwhile, the second criteria aims to keep the triangles that form the signal domain in the absence of strong interference between signal components, because all vertices of those triangles must be zeros in the border of the signal domain, i.e. zeros of the third kind. Given that these criteria are not based on the length of the triangles edges, they would allow to more efficiently separate the triangles corresponding to the signal from those corresponding to noise for the signal defined in Eq. \eqref{eq::triple_tone}. Results of this approach are described in the following section.

\section{Numerical Experiments} \label{sec::results}

In this section, we describe a series of numerical simulations that exemplify the use of the proposed classification into three kinds of the zeros of the spectrogram. First, we study the behavior of Algorithm \ref{alg::discriminating_zeros} for different parameter choices. Then, the results of the classification are showcased using synthetic and real signals as examples. Finally, the results of the proposed denoising strategy based on the classification of the zeros of the spectrogram are shown for the signal defined in Eq. \eqref{eq::triple_tone}.

\subsection{Setting of Parameters}\label{sec::results::parameters}

\begin{table}
\centering
  \setlength{\tabcolsep}{3pt}
 \renewcommand{\arraystretch}{1.2}
\caption{Studied values of the parameters of Algorithm \ref{alg::discriminating_zeros}. $J$ is the number of noise realizations used to build the 2D histograms of zeros. $r$ is the radius of the neighborhood of each original zero, expressed in terms of the width parameter $T$ of the analysis window.} \label{table::parameters}
\begin{tabular}{|c|c|} \hline
  Parameter & Values   \\ \hline
  $J$ & $64$, $128, 256, 512, 1024$ \\
  $r$ & $T/8, T/4, 3T/8, T/2, 5T/8$  \\ \hline
%   $\beta$ & $0.7,0.85,1.00,1.15,1.30$ \\ \hline
 \end{tabular}
\end{table}
 
Numerical simulations were carried out to explore the dependence of Algorithm \ref{alg::discriminating_zeros} on the number $J$ of noise realizations used to compute the 2D histograms, and the radius $r$ used to determine the neighborhood of the original zeros. The accuracy of the classification of zeros by means of Algorithm \ref{alg::discriminating_zeros}, i.e. the proportion of correctly classified zeros, was computed for a discrete set of values of $r$ and $J$, detailed in Table \ref{table::parameters}. With this purpose, a \emph{ground truth} set of correctly classified zeros according to the types described in Sec. \ref{sec::three_types} was generated, making use of the known frequencies of the three tones of the signal defined in Eq. \eqref{eq::triple_tone}. 

Then, optimal values of $r$ and $J$ were found as a solution of 
\begin{equation} \label{eq::argmax_accuracy}
 \argmax\limits_{r,J} \operatorname{Acc}(r,J),
\end{equation}
where $\operatorname{Acc}(r,J)$ is the accuracy of Algorithm \ref{alg::discriminating_zeros}, for the values of $r$ and $J$ given in Table \ref{table::parameters}. 

\begin{figure}
\centering
 \includegraphics{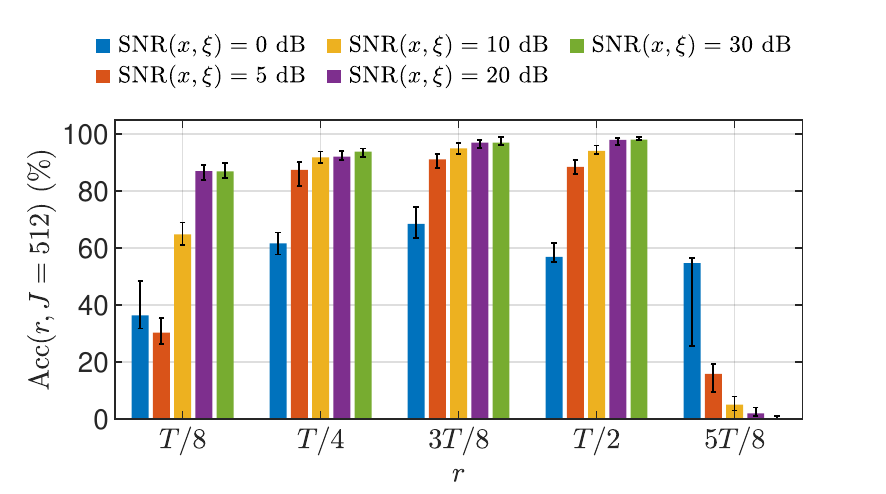}
\caption{Accuracy of the classification of zeros, i.e. the proportion of correctly classified zeros, for the signal defined in Eq. \eqref{eq::triple_tone} with $256$ time samples, using Algorithm \ref{alg::discriminating_zeros} with $J=512$ and different values of $r$. The parameter $T$ is the analysis window width in Eq. \eqref{eq::window}. The bars correspond to five values of $\operatorname{SNR}(x,\xi)$ and indicate the median of $100$ repetitions of the experiment. The lower (resp. upper) error bar denotes the $0.25$ (resp. $0.75$) quantile.} \label{fig::parameters} 
\end{figure}

Based on the simulations done, higher values of $J$ lead to a higher accuracy, but increasing $J$ from $512$ to $1024$ did not lead to a significant improvement of the classification.
Fig. \ref{fig::parameters} shows the accuracy of the classification of zeros for several values of the radius $r$, this latter being given in terms of the parameter $T$ of the analysis windows, and for $J=512$. For each value of $r$, the accuracy for five different noise conditions given by $\operatorname{SNR}(x,\xi)$ are shown, corresponding to the height of the bars. Each bar represents the median of $100$ estimations of the accuracy. Algorithm \ref{alg::discriminating_zeros} is highly sensible to the choice of $r$, in particular for lower SNR. Overall, $r=3T/8$ seems to maximize the accuracy for all the choices of SNR, although $r=T/4$ and $r=3T/8$ result in a similar performance for higher SNRs. Considering these results, in the following we will use $r = 3T/8$ and $J=512$, unless stated otherwise. 

\subsection{Classification of the Spectrogram Zeros}
Figs. \ref{fig::zeros_classification}a and \ref{fig::zeros_classification}b show the spectrograms of two different signals with two components and their classified zeros superimposed.
In the case of the two parallel linear chirps (Fig. \ref{fig::zeros_classification}a), it is clearly depicted how the different types of zeros are classified based on the proposed descriptors of the 2D histogram. The algorithm is also able to identify the zeros of the first kind created when a component interferes with \emph{itself} due to rapid changes in its instantaneous frequency, as it happens with the sinusoidal chirp shown in Fig. \ref{fig::zeros_classification}b.

Fig. \ref{fig::zeros_classification}c shows the result of applying Algorithm \ref{alg::discriminating_zeros} to a signal whose spectrogram has only zeros of the second and third kind. In such case, as expected, only two clusters are identified by Algorithm \ref{alg::discriminating_zeros}.

An example using a bat echolocation call signal is shown in Fig. \ref{fig::batsig_zeros}, where the log-spectrogram with the classified zeros superimposed and the 2D histogram are shown. Two things are noteworthy in this case. First, the exponential-like components are closer in the rightmost part of the signal. Thus, zeros of the first kind can be seen in that region, marked as blue dots. Another aspect is that the signal suffers from \emph{aliasing}, as it can be seen in the upper border of the spectrogram, where the components are \emph{reflected} and then generate interference with the legitimate components of the signal. This interference between the reflected components and the true ones also creates zeros of the first kind.

When the analyzed signal is made of noise only, then all the zeros of the spectrogram belong to the second kind. In this case, one would expect that Algorithm \ref{alg::discriminating_zeros} finds only one cluster ($K=1$) in the domain of the descriptors, whereas if a signal is present $K$ should be greater than $1$. This behavior is tantamount to a \emph{detection} method. In order to assess its detection performance, we applied the Algorithm \ref{alg::discriminating_zeros} to $200$ signals consisting of only real white Gaussian noise, and $200$ mixtures of a linear chirp (shown in Fig. \ref{fig::zeros_classification}c) and noise with different SNRs. Table \ref{table::detection_perf} summarizes the proportion of times $K=1$ was found when the signal is pure noise (specificity) and the proportion of times the algorithm indicated $K>1$ for a signal mixture (sensitivity). These results were compared with the detection tests proposed in Bardenet et al \cite{bardenet2018zeros, baddeley2014tests}, the performance of which is slightly better, as shown also in Table \ref{table::detection_perf}. 

\begin{table}  
  \setlength{\tabcolsep}{3pt}
 \renewcommand{\arraystretch}{1.3}
 \caption{Specificity and sensitivity on $200$ signals, using Algorithm \ref{alg::discriminating_zeros} (detecting if $K=1$ or $K>1$, with $r=3T/8$ and $J = 1024$) and the detection tests described in \cite{bardenet2018zeros,baddeley2014tests} (fixing the test significance to $0.14$ for a fair comparison). Clopper-Pearson $95\%$ confidence intervals are given in square brackets. Values in dB correspond to $\operatorname{SNR}(x,\xi)$.} \label{table::detection_perf}
\begin{center}
 \begin{tabular}{|c|c|c|c|c|} \hline
 \multirow{2}{*}{Method}    &\multirow{2}{*}{Specificity}   & \multicolumn{3}{c|}{Sensitivity}      \\ \cline{3-5}
                            &     & $0$ dB    & $5$ dB    & $10$ dB       \\ \hline
     Algorithm \ref{alg::discriminating_zeros}  & $0.86$ &    $0.51\;[0.43;0.58]$ & $0.90\;[0.85;0.94]$      &     $1.00$   \\ \hline
     Bardenet et al. \cite{bardenet2018zeros}  & $0.86^{\dagger}$     &    $0.59\;[0.51;0.65]$ & $0.99\;[0.97;1.00]$      &     $1.00$    \\ \hline
     \end{tabular}
     
 \end{center}
    {\raggedleft $\dagger$ Fixed by design.}
\end{table}

\begin{figure*}
\centering
 \setlength{\tabcolsep}{0pt}
 \renewcommand{\arraystretch}{0.5}
\begin{tabular}{ccc}
\includegraphics{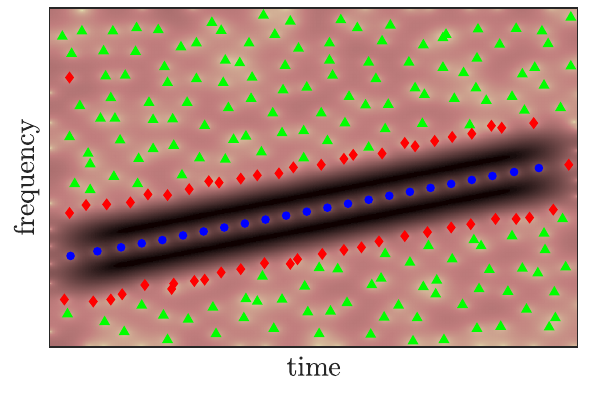} &
% (a)\\ 
 \includegraphics{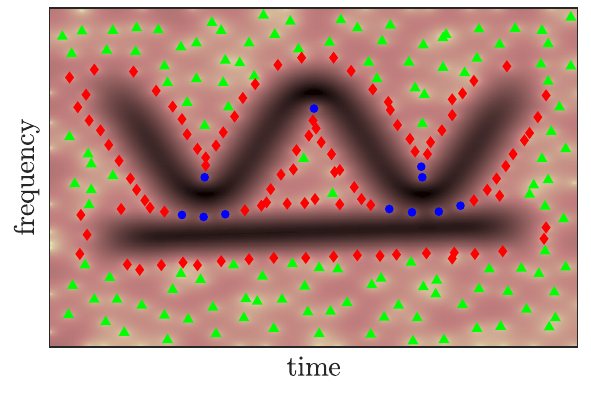} &
% (b) \\
 \includegraphics{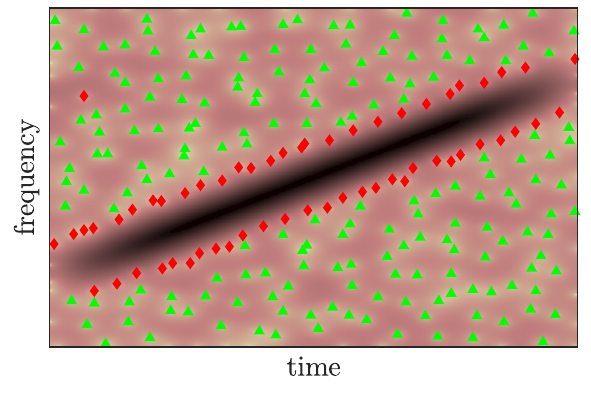} \\
% (c) {one_chirp}

(a) & (b) & (c) \\
\end{tabular}

\caption{Panels (a) and (b) show the spectrograms of different two-component signals with $\operatorname{SNR}(x,\xi)=30$ dB. Panel (c) shows the classification when only two types of zeros are present for a monocomponent signal. The values $r = 3T/8$ and $J=512$ were used here for the classification of zeros by means of Algorithm \ref{alg::discriminating_zeros}. Blue dots denote the zeros of the first kind, created between the components of the signal. Green triangles mark the position of zeros of the second kind, in the noise-only regions. Red diamonds represent the zeros of the third kind, created at the border of the signal domain due to interference between the signal and noise.} \label{fig::zeros_classification}
\end{figure*}

\begin{figure}
\centering
 \setlength{\tabcolsep}{0pt}
 \renewcommand{\arraystretch}{0}
  \hspace*{-0.38cm}
\begin{tabular}{cc}
  \includegraphics[width = 0.25\textwidth]{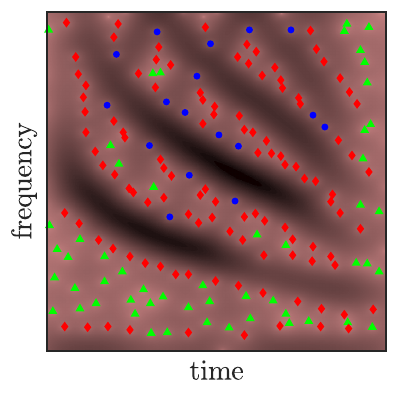} &
 \includegraphics[width = 0.25\textwidth]{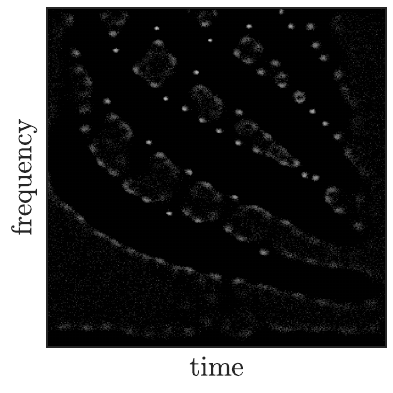} \\
 (a)& (b) \\

\end{tabular}
\caption{(a) Log-spectrogram of a bat echolocation signal with classified zeros superimposed. (b) The corresponding 2D histogram of the position of new zeros. The values $r = 3T/8$ and $J=512$ were used here for the classification of zeros by means of Algorithm \ref{alg::discriminating_zeros}. Blue dots denote the zeros of the first kind, created between the components of the signal. Green triangles mark the position of zeros of the second kind, in the noise-only regions. Red diamonds represent the zeros of the third kind, created at the border of the signal domain due to interference between the signal and noise.} \label{fig::batsig_zeros}
\end{figure}

\subsection{Signal and Noise Disentanglement}

\begin{figure*}
% \centering
%  \setlength{\tabcolsep}{0pt}
%  \renewcommand{\arraystretch}{0.5}
\begin{tabular}{ccc}
\centering
 \includegraphics[width=0.3\textwidth]{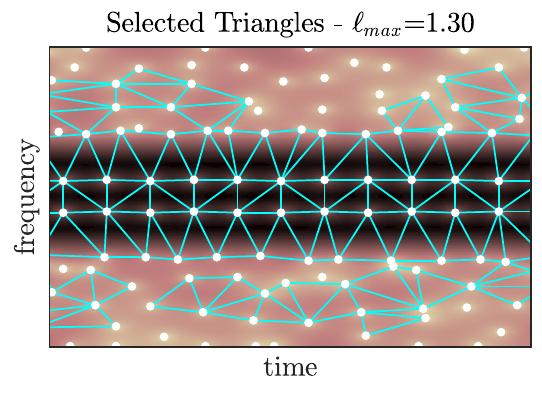} &
 \includegraphics[width=0.3\textwidth]{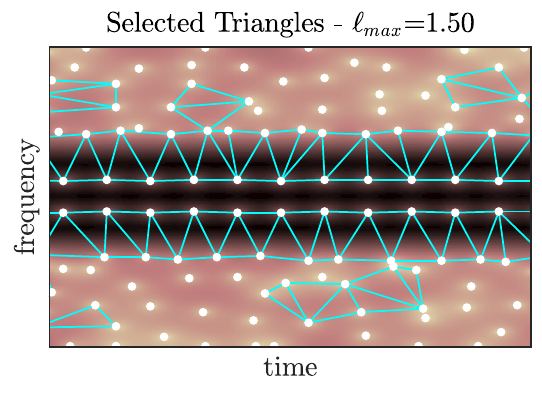} &
 \includegraphics[width=0.3\textwidth]{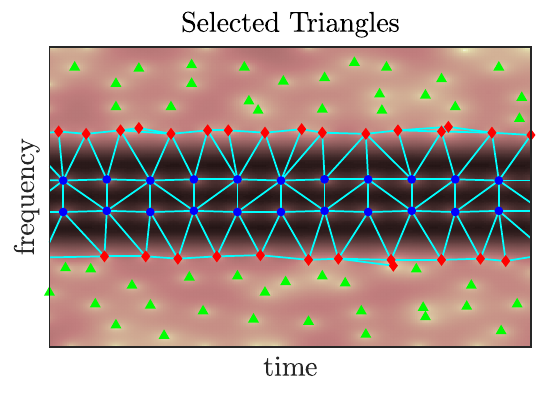} \\
 (a)& (b) & (c) \\
\end{tabular}
\caption{The three panels show the spectrogram of the signal described in Eq. \eqref{eq::triple_tone}, with the spectrogram zeros and the selected triangles superimposed. (a) Triangles selected by applying a threshold of $\ell_{\max}=1.30$. (b) Triangles selected by applying a threshold of $\ell_{\max}=1.50$. (c) Triangles selected after classifying the zeros using Algorithm \ref{alg::discriminating_zeros} and applying the criteria described in Sec. \ref{sec::tf_filtering}. Blue dots denote the zeros of the first kind, created between the components of the signal. Green triangles mark the position of zeros of the second kind, in the noise-only regions. Red diamonds represent the zeros of the third kind, created at the border of the signal domain due to interference between the signal and noise.}\label{fig::triangles}
\end{figure*}

\begin{figure}
\centering
 \includegraphics{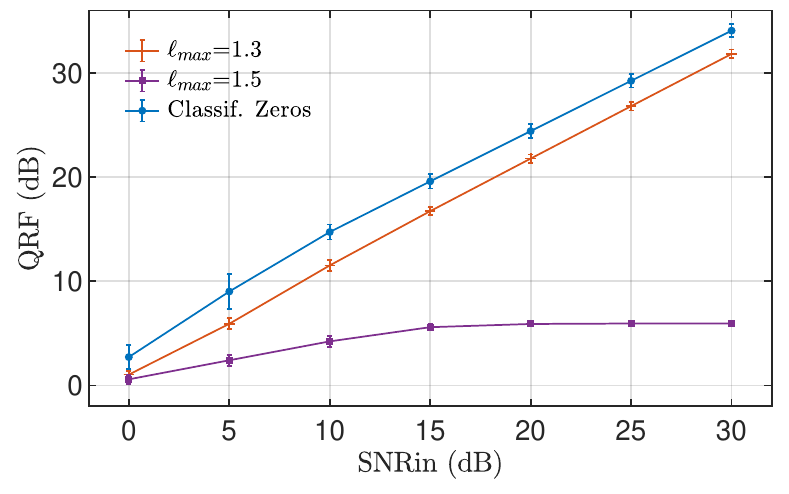} 
\caption{Comparing the performance of filtering approaches for the signal defined in in Eq. \eqref{eq::triple_tone}. ``Classif. Zeros'' is the proposed strategy of selecting triangles based on the classification of zeros using Algorithm \ref{alg::discriminating_zeros} with $r = 3T/8$ and $J=512$. The other curves correspond to the selection of triangles based on a threshold $\ell_{\max}$ (shown in the legend). Each marker indicates the mean over 200 repetitions of the experiment. The errorbars indicate the standard deviation.} \label{fig::snr_comparison}
\end{figure}

Here we compare the performance of the denoising strategy proposed in Sec. \ref{sec::tf_filtering}, with the approach proposed in \cite{flandrin2015time} using $\ell_{\max}=1.3$ and $\ell_{\max}=1.5$. The selected Delaunay triangles for these two thresholds are shown in Figs. \ref{fig::triangles}a and \ref{fig::triangles}b, respectively. It can be seen that with a larger threshold than the length computed using Eq. \eqref{eq::sep_condition}, a number of triangles corresponding to the signal domain are lost. On the other hand, when $\ell_{\max}=1.3$ one can find all the triangles inside the signal domain, but more triangles corresponding to noise are selected. In contrast to these situations, Fig. \ref{fig::triangles}c shows the triangles selected by first classifying the zeros and then choosing only the triangles that satisfy the mentioned criteria. Notice that we are not interested here in component reconstruction, but in correctly determining the signal domain. A systematic comparison between the two filtering strategies can be seen in Fig. \ref{fig::snr_comparison} using the Quality Reconstruction Factor (QRF) as a performance metric
\begin{equation}
 \operatorname{QRF} \coloneqq 10 \log_{10}\left( \frac{\norm{x}^2_{2}}{\norm{x-\tilde{x}}^{2}_{2}} \right)\; {\rm (dB)},
\end{equation}
where $x$ is the signal defined in Eq. \eqref{eq::triple_tone} and $\tilde{x}$ is the denoised signal obtained after classifying the zeros of the spectrogram using Algorithm \ref{alg::discriminating_zeros}, approximate the signal domain by selecting the triangles according to the criteria detailed in Sec. \ref{sec::tf_filtering}, and synthesize the signal from the STFT using the inversion formula, as done in \cite{flandrin2015time}.

\section{Discussion} \label{sec::discussion}
As exemplified in Fig. \ref{fig::clusters}, the measures of local density and concentration of the histogram cannot split sufficiently well all the (potentially) three clusters when used individually, which is why their combination is essential to correctly differentiate the clusters when the three types of zeros are present. We explored a number of possible local density and local concentration measures, like the 1-norm, 2-norm, the mean and other statistics on the number of zeros fallen inside the neighborhoods defined in Eq. \eqref{eq::vicinity}, as well as different measures of entropy for $\alpha$ in Eq. \eqref{eq::renyiH} (min-entropy for $\alpha\rightarrow\infty$, Shannon entropy for $\alpha\rightarrow 1$). We have chosen the measures that maximized the performance of the proposed approach from the ones that we have explored, yet we do not claim that the here used features ($\Vert B({\bf{z}},r)\Vert_{1}$ and $H_{1}(B({\bf{z}},r))$) are \emph{optimal}. Thus, other combinations of descriptors could be used with the same purpose and, perhaps, better performance.

We make here some remarks regarding the parameters of the method. Notice that, in average, the distance between the spectrogram zeros in the noise only case is approximately $T$ \cite{flandrin2015time, flandrin2016sound, flandrin2018explorations, bardenet2018zeros}. Hence, taking $r>T/2$ greatly increases the superposition of the balls defined in Eq. \eqref{eq::vicinity} which decreases the performance of Algorithm \ref{alg::discriminating_zeros}, as shown in Figure \ref{fig::parameters}.
In practice, the number of TF points inside the neighborhoods defined in Eq. \eqref{eq::vicinity} increases with the length of the signal. Because the entropy estimation depends on how well the 2D histogram can approximate the underlying distribution, increasing the length of the signal would also require increasing $J$.

Considering the three-tone signal defined in Eq. \eqref{eq::triple_tone}, one can find a condition to completely disentangle noise from the signal expressed in terms of the separation between adjacent tones $\theta$ in Eq. \eqref{eq::sep_condition}. For $\ell_{\max}=2.0$ as in \cite{flandrin2015time}, one finds that $\theta\approx 4.84$ provides triangles with sufficiently long edges so as to be effectively separated from noise. This signal is a simple example of a limitation of the algorithm proposed in \cite{flandrin2015time}, and illustrates the fact that the triangle \emph{elongation} described in \cite{flandrin2015time} can be limited by the presence of adjacent components of the signal. In contrast, the classification of zeros yields a different, more effective criterion for the selection of the associated triangles in this case.

As mentioned before, deciding if the number of clusters is one or more is akin to a signal detection test. However, sensitivity for adverse noise conditions (like $0$ dB) and specificity are affected by the ability of the algorithm to correctly estimate the number of clusters, as one can see from Table \ref{table::detection_perf}. At lower levels of noise, one could instead apply first more sensitive detection tests, such as the ones introduced in \cite{bardenet2018zeros}, to determine the presence of a signal prior to the classification of zeros and noise filtering. In that case, Algorithm \ref{alg::discriminating_zeros} can be slightly modified to only evaluate if the optimal number of clusters is two or three, and then proceed with the cluster labeling.

On the other hand, detecting whether the number of clusters is two or three amounts to identify the presence of strong interference between the components of the signal for a given analysis window. Such an interference, which in the limit case of two components very close to each other produces the so-called \emph{time-frequency bubbles} \cite{delprat1992asymptotic,delprat1997global}, has been shown to be a hindrance in ridge detection and instantaneous frequency estimation. Future work will continue to explore other uses of the classification of the spectrogram zeros like the detection of time-frequency bubbles and proper selection of the analysis window to avoid strong interference between modes. 

The 2D histograms introduced in this work are interesting objects by themselves, and another aspect to study in the future is its underlying probability distribution, which might explain the differences in the local density and concentration among the types of zeros. The empirical approach taken here avoids dealing with an appropriate model of this distribution, with the cost of needing several noise realizations to compute the histograms. Other approaches, like smoothing the 2D histogram using a low-pass 2D filter, could be used to improve the density estimation while avoiding the computational expense of increasing the number of noise realizations. Subsequent work will hence focus on studying and modeling these histograms, with the purpose of reducing the dependence on key parameters like $J$ or $r$ in order to classify the zeros.

\section{Conclusion} \label{sec::conclusions}
Focusing on the destructive interference between components in the STFT that gives birth to the zeros of the spectrogram, a classification scheme of these zeros in three categories was introduced along with a noise-assisted algorithm to automatically classify them in the proposed types. Additionally, a denoising strategy, based on the classification of zeros, was described. Numerical experiments showed that the proposed algorithm is able to discriminate between the three kinds of zeros of the spectrogram for synthetic and real signals. 
Moreover, it was shown that the approach introduced for noise disentanglement, which provides a different criteria to identify the Delaunay triangles that form the signal domain, can produce better results than previously proposed methods when the components of the signal are very close to each other, thus limiting the expansion of the signal domain between the components.

The code used this article, as well as supplementary material, can be found in: \url{https://github.com/jmiramont/spectrogram-zeros-classification}.

\section*{Acknowledgment}
The authors wish to thank Curtis Condon, Ken White, and Al Feng of the Beckman Institute of the University of Illinois for the bat data in Fig. \ref{fig::batsig_zeros} and for permission to use it in this paper.

\section*{Appendix. Proof of Proposition \ref{prop::conditions}} \label{appendix::1}

% \subsection{Proof of Proposition \ref{prop::conditions}}
Let us express the spectrogram of the mixture of two signals $x_{1}(t)$ and $x_{2}(t)$ as 
\begin{multline}
 S^{g}_{x_{1}+x_{2}}(t,f) = M_{x_{1}}(t,f)^{2} + M_{x_{2}}(t,f)^{2} \\ + 2 M_{x_{1}}(t,f)M_{x_{2}}(t,f) \cos\left(\Delta \Phi(t,f) \right)
\end{multline}
where $\Delta \Phi(t,f) = \Phi_{x_{2}}-\Phi_{x_{1}}$, i.e. the difference of the phase functions of the signals $x_{2}$ and $x_{1}$ respectively.

Then one can see that conditions are sufficient by directly replacing $\Delta \Phi(t,f) = (2n+1) \pi,\; n\in\mathbb{Z}$ obtaining
\begin{equation}
 S^{g}_{x_{1}+x_{2}}(t,f) = (M_{x_{1}}(t,f) - M_{x_{2}}(t,f))^{2},
\end{equation}
from where it follows that $S^{g}_{x_{1}+x_{2}}(t,f)=0$ if $M_{x_{1}}(t,f) = M_{x_{2}}(t,f)$.

To proof that the conditions are necessary, one can write $S^{g}_{x_{1}+x_{2}}(t,f)$ as
\begin{equation}
 S^{g}_{x_{1}+x_{2}}(t,f) = \alpha^2 + b\alpha + c = 0
\end{equation}
where $\alpha = M_{x_1}(t,f)$, $b = 2\cos(\Delta\Phi(t,f))M_{x_2}(t,f)$ and $c=M_{x_2}(t,f)^{2}$, and find the roots of  $S^{g}_{x_{1}+x_{2}}(t,f)$ as
\begin{equation}
 \alpha = - M_{x_2} \cos(\Delta\Phi) \pm M_{x_{2}} \sqrt{\cos(\Delta\Phi)^{2}-1}.
\end{equation}

Considering that $\alpha$ must be real and non-negative, it follows that $\Delta\Phi$ must be an odd multiple of $\pi$, which also means that $M_{x_1}=M_{x_2}$.

%-------------------------------------------------------------------------------------------------------------------------------

\ifCLASSOPTIONcaptionsoff
  \newpage
\fi



\bibliographystyle{IEEEtran}
\end{document}